\documentclass[prl,twocolumn,10pt,showpacs,letterpaper]{revtex4}
\usepackage{graphicx}
\begin{document}
\title[Strangelets as cosmic rays]{Reply to Comment on Strangelets as cosmic rays beyond the GZK-cutoff}
\author{Jes Madsen}
\affiliation{Department of Physics and Astronomy, University of Aarhus, DK-8000 \AA rhus C, Denmark}
\pacs{98.70.Sa, 12.38.Mh, 12.39.Ba, 24.85.+p}

%\begin{abstract}
%\end{abstract}
\date{October 8, 2003}
\maketitle

In \cite{ml} it was demonstrated that strangelets (stable lumps of quark
matter) have properties which circumvent both the acceleration problem
and the energy-loss problems facing more mundane candidates
for ultrahigh energy cosmic rays beyond
$10^{20}$ eV, such as protons and nuclei.

In the preceding Comment Balberg \cite{balberg} argues that 
such a galactic flux of ultrahigh energy strangelets would trigger 
transformation of all neutron stars into strange quark matter stars. 
He further argues, that all neutron stars can not be strange stars, 
and therefore finds the scenario in \cite{ml} unlikely.

Here I show that the first assumption in \cite{balberg} 
is incorrect because strangelets
at the relevant energies will be destroyed in collisions with the stars
they are supposed to transform. I further argue, that it is not
currently known whether all neutron stars are in fact strange stars.
Thus the scenario presented in \cite{ml} remains viable. 

Strangelet fragmentation will occur if
the total energy added in inelastic collisions with
nuclei exceeds the strangelet binding energy, which can be
some tens of MeV per baryon. 
A strangelet with baryon number $A$ 
will be destroyed in a single head-on collision with a
stellar proton if $E_{\rm col}= 10^{20}{\rm eV}A^{-1}E_{20}>E_{\rm bind}
\approx 10^7 {\rm eV} A E_{B10}$ or $A<3\times 10^6 E_{20}^{1/2}
E_{B10}^{-1/2}$, where $E_{B10}$ is the
binding energy per baryon in units of 10 MeV, and $E_{20}$ is the cosmic ray
kinetic energy in units of $10^{20}$ eV.
Bringing a strangelet to rest in the star requires 
the strangelet to move through a total column of mass
roughly equal to its own, i.e.\ of order $A$ collisions with protons
or a fragmentation limit up to $A<10^{13} E_{20} E_{B10}^{-1}$ 
(highly charged strangelets will 
have a scattering cross section larger than
geometrical, which can reduce this $A$-limit somewhat). 
Smaller strangelet fragments potentially formed in such
ultra-relativistic collisions will have the same Lorentz factor and
energy per baryon as the original strangelet, and will therefore be
prone to destruction in later collisions. While the details of
strangelet fragmentation remains to be studied (much of the necessary
input physics is poorly known) these order-of-magnitude estimates
demonstrate that the ultrahigh energy strangelets discussed in
\cite{ml} will be destroyed in collisions rather than serve as seeds to
transform neutron stars into strange stars.

In contrast to the extremely high energy cosmic ray strangelets,
even a small flux of strangelets at low energies 
would be able to convert all neutron stars
into strange stars. This was first shown in detail in \cite{madsen1988}
and \cite{frca} more than a decade ago. At that time it was argued
(also by the present author in \cite{madsen1988}) that this ruled out
the hypothesis of stable strange quark matter and strangelets, because
some properties of pulsars (notably the glitch phenomenon) seemed
inconsistent with strange star properties. Balberg \cite{balberg}
revives this argument and lists a set of such
properties including glitches, r-mode instabilities and cooling.
However at the current level of understanding it is premature to rule
out strange stars on these grounds. The strange star glitch problem
\cite{alpar} has been shown to be marginally consistent with ordinary
strange stars with nuclear crusts \cite{webglen}, and may also find an
explanation in the crystalline phases recently discovered in 
color-superconducting
quark matter \cite{alford}. The r-mode instabilities that rule
out the simplest color-flavor locked strange stars \cite{madsen00} may
also be consistent in models with crystalline phases,
and similar counterexamples exist for the other effects mentioned in
\cite{balberg}.
Summarizing, it is not known at present whether strange stars exist
at all, but
it is not ruled out either, that all ``neutron'' stars could be strange
stars.

In conclusion, the strangelet scenario for ultrahigh energy cosmic rays
presented in \cite{ml} remains viable.

This work was supported by the Theoretical Astrophysics Center under
the Danish National Research Foundation, and by the Danish Natural Science
Research Council.

\end{document}